\RequirePackage{etex}                              
\documentclass{aa} 
\usepackage{natbib}
\bibpunct{(}{)}{;}{a}{}{,}
\usepackage{graphicx}
\usepackage{etex}
\usepackage{txfonts}
\usepackage{lipsum}
\usepackage{adjustbox}
\usepackage{subcaption}         
\usepackage{lscape}                                           
\usepackage{placeins}                                          
\usepackage{xcolor}
\usepackage{arydshln}
\usepackage{hhline}

\begin{document}

   \title{Time lags as proxy of spectral evolution in gamma-ray bursts}

   \author{C. Maraventano
          \inst{1,2}
    \and 
          F. Daigne
          \inst{3, 4}
    \and 
          R. Mochkovitch
          \inst{3}
    \and 
          L. Nava 
          \inst{2, 6}     
    \and 
          G. Ghirlanda
          \inst{2, 5}
    \and
          T. Di Salvo 
          \inst{1}}

   \institute{Dipartimento di Fisica e Chimica - Emilio Segrè, 
              Università di Palermo, via Archirafi 36 - 90123 Palermo, Italy
         \and
             INAF - Osservatorio Astronomica di Brera, Via E. Bianchi 46, I-23807, Merate (LC), Italy
        \and 
            Sorbonne Université, CNRS, UMR 7095, Institut d’Astrophysique de Paris, 98 Bis bd Arago, 75014 Paris, France
        \and 
            Institut Universitaire de France, Ministère de l’Enseignement Supérieur et de la Recherche, 1 rue Descartes, 75231 Paris Cedex F-05, France
        \and 
            INFN – Sezione di Milano–Bicocca, Piazza della Scienza 3, 20126 Milano (MI), Italy
        \and 
            INFN, Sezione di Trieste, I-34127 Trieste, Italy}
   \date{}

\abstract
   {Positive lags in gamma-ray bursts (GRBs) provide a unique window into the temporal evolution of their prompt emission, where hard photons anticipate softer ones. Negative lags, when hard photons are delayed, are instead more enigmatic to interpret. Disentangling the different effects that produce both kinds of lags is therefore critical for identifying the physical mechanisms at work in the prompt and early afterglow phases of GRBs.}
   {We investigate the potential of time lags for distinguishing the presence of different emission components at different energy bands. Considering data from the \emph{Fermi} Gamma-ray Burst Monitor (GBM) and the LAT Low Energy (LLE) technique, we aim at establishing a connection between lag behavior and high-energy spectral properties.}
   {We perform a time-resolved joint spectral analysis in the range 10~keV--100~MeV for two exceptionally bright bursts, GRB~160625B and GRB~190114C. Time lags between the lowest-energy band (10--100~keV) and progressively higher-energy bands up to 30--100~MeV were computed across their distinct emission episodes by means of the cross-correlation function.}
   {For GRB~160625B, the spectra are described by a single component with clear hard-to-soft evolution, and the time lags are always positive. Analysis of the high-energy exponential cutoff, likely originating above the photosphere, yields bulk Lorentz factor estimates of $\Gamma \sim 120-250$. GRB~190114C exhibits negative lags in the 30--100~MeV band, coinciding with a delayed additional high-energy powerlaw component that begins to dominate the LLE range after $\sim$2.5~s. Comparison with multi-wavelength observations shows some compatibility with the early afterglow, though its physical origin remains an open question, leaving room for interpretations such as external shocks or internal dissipation.}
   {Time lags are effective diagnostic tools to investigate the spectral evolution of GRBs. Positive lags trace the softening of the prompt emission, whereas negative lags, although more difficult to interpret, indicate the appearance of a new, independent high-energy spectral component.}

   \keywords{gamma-ray burst: general – radiation mechanisms: non-thermal – gamma-ray burst: individual: GRB 160625B, GRB 190114C}

   \maketitle
   \nolinenumbers

\abstract

\section{Introduction} \label{introduction} 

Gamma-ray bursts (GRBs) are short-lived flashes of $\gamma$-ray radiation originating at cosmological distances and releasing up to $\sim10^{54}$ erg in isotropic equivalent energy. They are characterized by two distinct emission phases: an initial, highly variable prompt emission of hard X-rays and $\gamma$-rays, followed by a multi-wavelength afterglow decaying smoothly from Very High Energy (VHE, $>$\,100\,GeV) to radio frequencies (see \citealt{kumar2015physics, nava2021gamma}). GRBs are historically classified into long ($>$2 s) and short ($<$2 s) based on their prompt duration, linked to the collapse of massive stars and the merger of compact binaries, respectively\footnote{The recent discovery of long GRBs with a compact binary progenitor \citep{zhong2023grb, bulla2023grb} and vice versa \citep{rossi2022peculiar} is increasingly challenging this traditional classification.}.

The physical mechanisms responsible for the prompt emission remain highly uncertain. In the internal shock model, collisions between shells of different Lorentz factors within an ultra-relativistic outflow produce non-thermal radiation, accounting for the observed fast temporal variability \citep{rees1994unsteady, kobayashi1997can, daigne1998gamma}. Alternatively, magnetic reconnection models invoke dissipation through reconnection events at large radii in highly magnetized outflows \citep{zhang2011internal, mckinney2012reconnection, beniamini2016properties}. Both models predict prompt emission originating above the photosphere, consistent with the observed non-thermal spectra. However, photospheric models have shown that dissipation near the photosphere can broaden a thermal spectrum into a non-thermal one, resembling the observed prompt spectra \citep{rees2005dissipative, pe2017photospheric, beloborodov2017photospheric}. Moreover, the possible presence of an extra component besides the usual prompt emission \citep{zhang2011comprehensive} adds further complexity to this scenario, making it crucial to identify and separate these contributions.

Time lags (or spectral lags), defined as the time delay between photons of different energies, and measured via cross-correlation of GRB light curves \citep{band1997gamma, peterson1998uncertainties}, are found to be a common feature of long GRBs, with magnitudes ranging from fractions of a second to tens of seconds \citep{norris2005long, chen2005distribution, ukwatta2010spectral}. Data from the Burst And Transient Source Experiment (BATSE) on board the Compton Gamma Ray Observatory (CGRO) showed that most long GRBs exhibit positive lags, with lower-energy photons arriving later, while only $\sim$10–20\% show negative lags \citep{hakkila2007gamma}. \emph{Swift}/BAT data confirmed positive or near-zero lags for long GRBs in the hard X-ray band \citep{ukwatta2010spectral, bernardini2015comparing}, whereas short GRBs exhibit negligible lags \citep{norris2006short, zhang2006revisiting, berger2014short, bernardini2015comparing}. Above $\sim$\,100\,MeV, observations with the \emph{Fermi} Large Area Telescope (LAT) revealed a few negative lags between the MeV and GeV emission, either in individual bright bursts or in systematic studies of LAT-detected GRBs \citep{ackermann2013multiwavelength, castignani2014time, ajello2019decade}. 

If a single emission component is present, positive lags are thought to arise from the hard-to-soft spectral evolution of the GRB spectrum \citep{kocevski2003connection, ryde2005interpretations, shao2017new}. Time lags behavior is indeed affected by the temporal evolution of both spectral and temporal parameters of the GRB during the propagation of the emitting region, such as its spectral indices, characteristic energies, shape and spikiness of the light curves \citep{daigne2003physics, bocci2010lag, mochkovitch2016simple}. Positive time lags may also be influenced by curvature effects associated with high-latitude emission from the relativistically-expanding shell \citep{peng2011spectral, uhm2016evidence}. Negative lags are more complex to interpret: they have been proposed to signal the delayed emergence of an additional high-energy spectral component \citep{ackermann2013multiwavelength, ajello2019decade}, identified either as onset of the external afterglow \citep{gupta2021grb} or as Inverse Compton emission from internal shocks \citep{chakrabarti2018spectral, bovsnjak2009prompt, bovsnjak2014spectral}. Time lags can therefore encode both the spectral shape and its temporal evolution, making them a powerful diagnostic to probe GRB emission mechanisms, particularly when analyzed over a broad energy range.

Extracting meaningful information from time lags requires simultaneous observations over distinct, non-overlapping energy bands, a capability enabled by the \emph{Fermi} Gamma-ray Space Telescope. Through the synergy of the Gamma-ray Burst Monitor (GBM; \citealt{meegan2009fermi}) and the LAT \citep{atwood2009large}, continuous data coverage from 8~keV to over 100~GeV is provided. While the standard LAT data-analysis considers photons with energies $>$ 100 MeV, the LAT Low Energy (LLE) technique is a more recent analysis which has been developed to reconstruct photon energies from 30 to 100 MeV \citep{pelassa2010lat}. These data fill the gap between the GBM and the standard LAT energy ranges, thus allowing to disentangle, if present, the prompt emission from the early afterglow component. The GeV emission detected by LAT has, in several cases, been interpreted as early afterglow radiation \citep{ghisellini10,ghirlanda2010onset, abdo2009fermi_second, kumar2010external, ajello2019decade}.  

In \cite{maraventano2025high}, hereafter M25, the systematic analysis of 70 GRBs showed that time lags computed between the GBM-NaI (10--100~keV) and LLE (30--100~MeV) bands are positive in the 40\% of bursts and negative in 37\% of them, while the four short GRBs in the sample showed negligible lags. These results were interpreted within a simple phenomenological scenario, in which positive lags arise due to the hard-to-soft spectral evolution of the prompt emission, while negative lags may be produced by the emergence of a distinct high-energy spectral component. However, this work presented only a time-integrated analysis of LLE spectra, so that it remains unclear whether positive lags reflect the hard-to-soft spectral evolution of the prompt emission, and negative lags are truly associated with the delayed emergence of a high-energy component. 

In this work, we investigate the potential of time lags as a complementary diagnostic tool for probing the emission mechanisms in GRBs, particularly in situations where detailed spectral modeling is limited by poor photon statistics or data quality. Extending the analysis of M25, we perform a time-resolved joint GBM+LLE spectral study of two exceptionally bright \emph{Fermi} bursts, GRBs~160625B and 190114C, selected for their high signal-to-noise ratio and broad energy coverage. These characteristics allow us to track the spectral evolution and time-lag behavior across their distinct emission episodes with high sensitivity. We then examine the connection between spectral evolution and time lags and test two physical scenarios: the hard-to-soft evolution of a single spectral component versus the delayed emergence of an independent high-energy component.

We describe the data selection and extraction in Section~\ref{methods}. The methods adopted for computing time lags and performing time-resolved spectral analysis are presented in Section~\ref{analysis}. We present the results in Section~\ref{results} and discuss them in Section~\ref{interpretation}. We give our conclusions in Section~\ref{discussion}. Throughout the analysis process, we adopted the standard $\Lambda$CDM cosmology of \citet{aghanim2020planck}.

\section{Data selection and extraction} \label{methods} 

In the following, we describe the criteria used to select the two GRBs analyzed, the extraction of their light curves and spectra from \emph{Fermi}-GBM and LLE data, and the strategy for selecting time intervals for the time-resolved spectral analysis and time lag computation.

\subsection{Selected GRBs} \label{selected_grbs} 

We selected two bright \emph{Fermi} GRBs: 160625B and 190114C. Both exhibit high LLE detection significance (49 and 53$\sigma$, respectively), as quantified by the Bayesian blocks algorithm \citep{ajello2019decade}, and high photon statistics at lower energies, with fluences of $6.43$ and $4.43 \times 10^{-4}\,\mathrm{erg\,cm^{-2}}$ in the 10--1000~keV band, corresponding to approximately the 85th and 76th percentiles of the GBM catalog fluence distribution \citep{von2020fourth}. Both events also display light curves with several bright, distinguishable pulses, enabling a detailed time-resolved joint GBM+LLE (10~keV--100~MeV) spectral analysis and time lag computation between the lowest (10--100~keV) and highest (30--100~MeV) energy ranges allowed from the data. 

\subsection{Extracting lightcurves and spectra}

For both GRBs, GBM Time-Tagged Event (TTE) files were retrieved from the \emph{Fermi} online archive\footnote{\url{https://heasarc.gsfc.nasa.gov/W3Browse/fermi/fermigbrst.html}} for the NaI and BGO detectors with the highest count rates. Photon events were binned at 0.1~s in three non-overlapping energy bands corresponding to the NaI (10--100~keV, Band~1), and BGO (150--500~keV, Band~2, and 500~keV--1~MeV, Band~3) detectors. For LAT/LLE data, TTE files were downloaded from the FERMILLE catalog\footnote{\url{https://heasarc.gsfc.nasa.gov/W3Browse/fermi/fermille.html}}, pre-selected following \citet{pelassa2010lat}, and filtered in the 30--100~MeV range (Band~4)\footnote{These four non-overlapping bands follow the same scheme as M25, ensuring adequate photon statistics while tracking time lag evolution with energy.}, also binned at 0.1~s. The background was estimated from off-source intervals before and after the GRB, fitted with polynomial functions up to the fourth order, with the best fit selected by lowest $\chi^2$ and extrapolated over the burst interval.
The lightcurves of GRBs~160625B and 190114C extracted in Band 1, 2, 3 and 4 are shown in panels A-D of Fig.~\ref{parameters_grb160625b} and Fig.~\ref{parameters_grb190114c}, respectively. 

GBM and LLE spectral data products, including the response matrix files (rsp2) were retrieved from the online archive. Given that both GRBs are long-duration events, we used CSPEC data, with a temporal resolution of 1024~ms. Spectra were generated with the \texttt{GTBURST}\footnote{\url{https://fermi.gsfc.nasa.gov/ssc/data/analysis/scitools/gtburst.html}} software, covering 10--900~keV for NaI, 0.3--10~MeV for BGO, and 30--100~MeV for LLE. Channels between 30 and 40~keV were excluded to avoid systematic effects from the Iodine K-edge at 33.17~keV\footnote{\url{https://fermi.gsfc.nasa.gov/ssc/data/analysis/GBM_caveats.html}}.

\subsection{Time intervals selection}

Time intervals for the time-resolved spectral analysis were selected in a way to ensure sufficient signal-to-noise across all detectors. We first considered intervals which separate the distinct pulses in each GRB. Since the LLE data have the lowest photon statistics, we then ensured that each interval contained at least 50 LLE counts (in the 30--100~MeV band) after background subtraction. The same binning scheme was used for the time lag computation. The selected intervals are listed in Tables~\ref{160625B_results_table} for GRB~160625B and Table \ref{190114C_results_table} for GRB~190114C. For GRB~160625B, the first pulse was split into its rise and decay sub-intervals, given its long duration and intensity (see Figure~\ref{parameters_grb160625b}, panels A--D).

\section{Analysis} \label{analysis} 
In this section we describe the spectral models and model-selection criterion for the joint GBM+LLE spectral analysis of GRBs~160625B and 190114C, as well as the time lag computation method applied to their distinct emission pulses.

\subsection{GBM + LLE spectral analysis} 

The joint GBM+LLE time-integrated and time-resolved spectral analyses were performed with \texttt{Xspec} (version 12.14.1), adopting the PG-statistic. We applied an inter-calibration factor among all detectors, normalized to the NaI one, and allowed it to vary within 30\%.

We modeled the spectra with a double smoothly broken power law (2SBPL; see \citealt{ravasio2018consistency} for its functional form), consisting of three power laws with photon indices $\alpha_1$, $\alpha_2$, and $\beta$ (from the lowest to the highest energies), connected at two break energies: $E_\mathrm{break}$ (separating the $\alpha_1$ and $\alpha_2$ power-law segments) and $E_\mathrm{peak}$ (separating the $\alpha_2$ and $\beta$ segments). Since typically $\alpha_2>-2$ and $\beta<-2$, $E_\mathrm{peak}$ represents the peak of the  $\nu F_\nu$ spectrum.

For GRB~160625B, we investigated the presence of a high-energy spectral cutoff (as found in \citealt{ravasio2024insights} for the \emph{Fermi} time-integrated spectrum from GBM to LAT energies) adopting the \texttt{highecut} multiplicative model from \texttt{Xspec}:

\begin{equation}
    \mathrm{highecut(E)} = \left\{ \begin{array}{cl}
1 & \mathrm{for} \ E \leq E_\mathrm{c} \\
e^{(E_\mathrm{c} - E)/E_\mathrm{fold}} & \mathrm{for} \ E \ge E_\mathrm{c}
\end{array} \right.
\end{equation}

where $E_\mathrm{c}$ is the cutoff onset energy and $E_\mathrm{fold}$ controls the sharpness of the transition. The model flux is reduced by a factor $1/e$ at $E_\mathrm{cutoff} = E_\mathrm{c} + E_\mathrm{fold}$, hereafter referred to as the cutoff energy. The resulting model (2SBPL $\times$ \texttt{highecut}) is denoted with 2SBPLCUTOFF.

For GRB~190114C, following \citet{ajello2020fermi}, we tested the presence of an additional high-energy component by adding a power law (PL) to the 2SBPL:

\begin{equation}
    \mathrm{PL}(E) = \mathrm{Norm_{PL}}\,E^{-\Gamma_\mathrm{PL}}
\end{equation}

where $\mathrm{Norm_{PL}}$ is the normalization in units of $\mathrm{photons\,keV^{-1}cm^{-2}s^{-1}}$ and $\Gamma_\mathrm{PL}$ is the photon index.\\

The best-fit model was chosen based on the Akaike Information Criterion (AIC; \citealt{1100705}): the more complex model (2SBPLCUTOFF or 2SBPL+PL) is preferred over the 2SBPL model when $\Delta \mathrm{AIC} \ge 6$ \citep{burnham2004multimodel}, where $\Delta \mathrm{AIC} = \mathrm{AIC}_i - \mathrm{AIC}_{\min}$ is the difference between the AIC value of the $i$-th model and the minimum AIC among the tested models\footnote{The relative likelihood of the $i$-th model is $e^{-\Delta\mathrm{AIC}/2}$; for $\Delta\mathrm{AIC}=6$, this corresponds to $\sim$0.05, meaning the $i$-th model is about twenty times less probable than the best-fit model.}. All spectral parameter uncertainties are reported at $1\sigma$ confidence level.

\subsection{Time lags computation in different pulses} 

We estimated time lags between the lowest-energy GBM-NaI band (Band~1; 10--100~keV) and the three progressively higher-energy bands, GBM-BGO Band~2 (150--500~keV), Band~3 (500~keV--1~MeV), and LAT-LLE Band~4 (30--100~MeV). The corresponding time lags were denoted as $\tau_{12}$, $\tau_{13}$, and $\tau_{14}$, respectively, and computed both over the full burst duration and over the same time intervals used for the time-resolved spectral analysis.

Following M25, time lags were computed via the discrete correlation function (DCF) method between pairs of background-subtracted light curves binned at 0.1~s. The lag $\tau$ is defined as the DCF maximum, determined by fitting the DCF with an asymmetric Gaussian, to account for the intrinsic asymmetry of GRB pulses, which allows to resolve lags smaller than the bin size. Uncertainties were estimated via a flux-randomization Monte Carlo technique \citep{peterson1998uncertainties}: for each energy band, $N=10000$ simulated light curves were generated by randomizing count rates according to their Poisson uncertainties, and the DCF was computed and fitted for each realization. The resulting $\tau$ distribution was fitted with a Gaussian, whose mean and standard deviation define the best-fit lag and its $1\sigma$ uncertainty. We note that this procedure accounts for statistical uncertainties only, and does not capture systematic effects arising from 
overlapping pulses in multi-pulse light curves, which may bias the DCF towards incorrect lag values. This limitation is most relevant for time-integrated estimates, and is mitigated by a time-resolved analysis, which isolates individual emission episodes. Furthermore, we caution that, in the presence of overlapping emission components, the DCF lag may reflect a combination of contributions from different spectral components; its interpretation therefore strongly relies on spectral analysis results, as performed in this work. 

\begin{figure*}
\sidecaption
  \includegraphics[width=12cm]{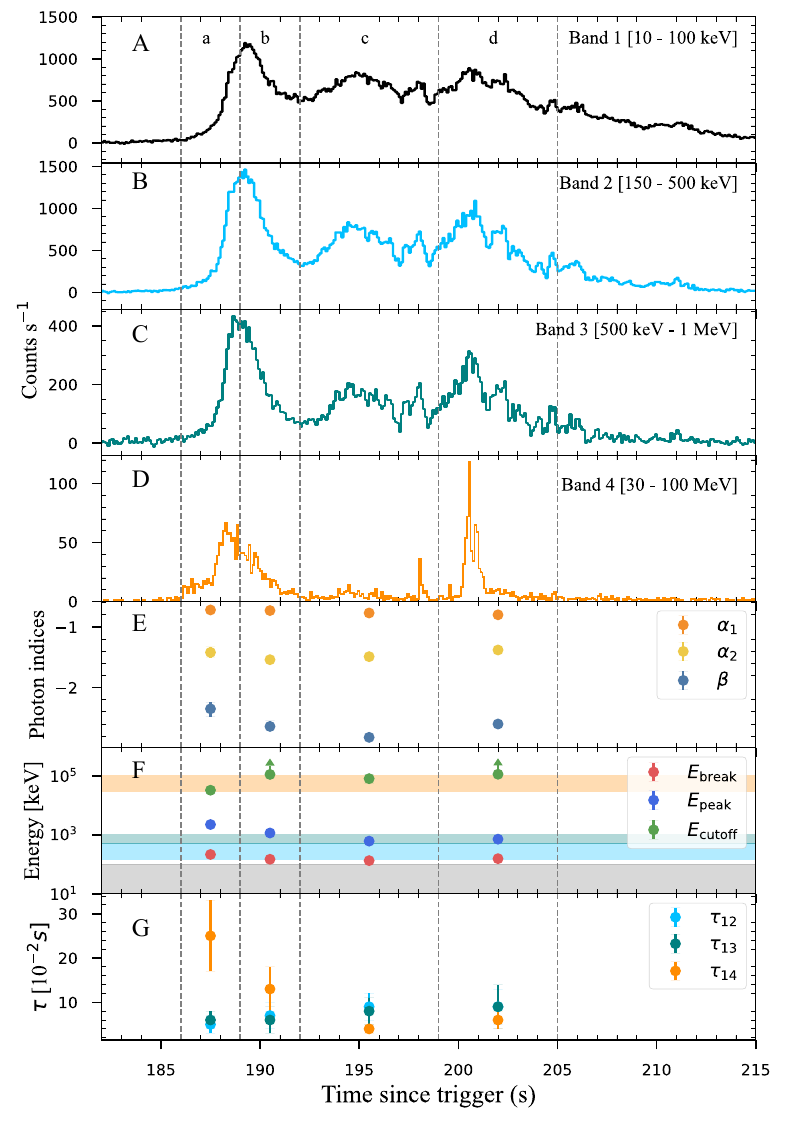}
     \caption{Time evolution of the spectral parameters and time lags obtained from the 2SBPLCUTOFF model (Table~\ref{160625B_results_table}) for the time-resolved spectra of GRB~160625B. Vertical gray dashed lines indicate the time intervals adopted for the spectral analysis. Panels A - D display the count-rate light curves in Bands~1, 2, 3, and~4, respectively, binned at 0.1~s. Panel E shows the photon indices below the break, between the break and the peak energy, and above the peak energy (orange, yellow, and blue symbols, respectively). Panel F presents the temporal evolution of $E_\mathrm{peak}$, $E_\mathrm{break}$, and $E_\mathrm{cutoff}$ (red, blue, and green symbols, respectively). The light-gray, blue, green, and orange shaded regions mark the energy ranges corresponding to Bands~1, 2, 3, and~4. Panel G shows the time lags between Band~1 and Bands~2, 3, and~4 (blue, green, and orange symbols, respectively; Table~\ref{time_lags}). Arrows indicate lower limits.}
     \label{parameters_grb160625b}
\end{figure*}

\section{Results} \label{results} 

Tables~\ref{160625B_results_table} and \ref{190114C_results_table} list the results of the joint GBM+LLE spectral analysis of GRBs~160625B and 190114C, respectively, including the time intervals, best-fit parameters, PG-statistic over degrees of freedom, and $\Delta\mathrm{AIC} = \mathrm{AIC_{COMPLEX}} - \mathrm{AIC_{2SBPL}}$, where $\mathrm{AIC_{COMPLEX}}$ refers to 2SBPLCUTOFF (GRB~160625B) or 2SBPL+PL (GRB~190114C) related fit. Table~\ref{time_lags} reports the time lag results for both GRBs, with the time bins and corresponding values of $\tau_{12}$, $\tau_{13}$, and $\tau_{14}$ with their uncertainties.

In the following, we discuss the spectral and time-lag evolution of each GRB, as well as the evidence for a high-energy cutoff in GRB~160625B and an additional power-law component in GRB~190114C. These results are then discussed in Section~\ref{interpretation}.

\subsection{GRB~160625B} \label{section_160625B} 

\subsubsection{Time-integrated spectral analysis} 

We fitted the time-integrated spectrum of GRB~160625B over 186--205~s interval from the trigger time\footnote{The precursor is excluded from this fit.}. The fit performed with the 2SBPLCUTOFF model yields $\Delta\mathrm{AIC} \sim 96$ with respect to the 2SBPL one, confirming the presence of a high-energy spectral cutoff (Figure~\ref{spectra_grb160625b}, top panel). In the following section we expand this analysis by performing a time resolved spectral analysis including \emph{Fermi}-LLE data. 

\subsubsection{Time-resolved spectral analysis} 

Following Section~\ref{methods}, we divided the entire emission episode of GRB~160625B (186--205~s) into four time bins (a--d; gray dashed lines in Figure~\ref{parameters_grb160625b}, panel A - D). Both the 2SBPL and 2SBPLCUTOFF model were fitted in each bin; the 2SBPLCUTOFF model is preferred in all bins ($\Delta\mathrm{AIC} > 6$), though only lower limits on $E_\mathrm{cutoff}$ are obtained in bins \emph{b} and \emph{d}. The best-fit models is shown in Figure~\ref{spectra_grb160625b} for each interval. 

Figure~\ref{parameters_grb160625b} shows the light curves (panels A--D, binned at 0.1~s in Bands 1--4) and the temporal evolution of the 2SBPLCUTOFF parameters. The photon indices $\alpha_1$ and $\alpha_2$ are approximately constant over time, with mean values $\bar{\alpha}_1 = -0.75 \pm 0.04$ and $\bar{\alpha}_2 = -1.46 \pm 0.07$, while $\beta$ decreases over the first three bins, and $\bar{\beta} = -2.60 \pm 0.19$. Although the compressed $y$-scale of panel F in Figure~\ref{parameters_grb160625b}  makes it difficult to appreciate the spectral evolution visually, both $E_\mathrm{peak}$ and $E_\mathrm{break}$ show a softening over time: $E_\mathrm{peak}$ decreases rapidly during the first $\sim$6~s and then more slowly, while $E_\mathrm{break}$ drops sharply in the first $\sim$6~s before stabilizing around $\sim$145~keV. The evolution of $E_\mathrm{cutoff}$ is less straightforward; the lower limits in bins \emph{b} and \emph{d} suggest the cutoff may lie above 100~MeV.

\subsubsection{Time-lag analysis} 

Panel G of Figure~\ref{parameters_grb160625b} shows the evolution of $\tau_{12}$, $\tau_{13}$, and $\tau_{14}$ for GRB~160625B. All time lags are positive, indicating that high-energy photons systematically precede lower-energy ones across the full 10~keV--100~MeV range.

The values of $\tau_{12}$ and $\tau_{13}$ are mutually consistent and of order $10^{-2}$~s, as expected given the proximity in energy of Bands~2 and~3 and their common BGO detector, resulting in a closely related temporal structure of their light curves. In addition, the sensitivity of the DCF method may limit our ability to resolve small variations for time lags of this magnitude. In contrast, $\tau_{14}$ is systematically larger, up to an order of magnitude in the earliest bins, and shows a slow decrease over time, suggesting an evolution of the dominant emission process or physical conditions of the emitting region at hundreds of keV to MeV energies. This behavior is interpreted in Section~\ref{interpretation} in the context of a single, softening emission component, identified as the prompt. We also note that the burst-integrated lag values are larger than those measured in the individual time bins (Table~\ref{time_lags}), likely because the DCF applied to the full, multi-peaked light curve correlates contributions from multiple overlapping pulses, inflating the resulting lag.

We also note that in intervals~\emph{c} and \emph{d}, $\tau_{14}$ decreases to values comparable to or below $\tau_{12}$ and $\tau_{13}$. To further investigate this, we explicitly computed the lags between Band~4 and Bands~2 and~3, $\tau_{24}$ and $\tau_{34}$, for these intervals. In bin~\emph{c}, we find $\tau_{24} = (-1 \pm 3) \times 10^{-2}$~s and $\tau_{34} = (-2 \pm 2) \times 10^{-2}$~s, while in bin~\emph{d}, $\tau_{24} = (-3 \pm 3) \times 10^{-2}$~s and $\tau_{34} = (-3 \pm 3) \times 10^{-2}$~s. While the central values of $\tau_{24}$ and $\tau_{34}$ are negative, they are all compatible with zero within $1\sigma$, and no statistically significant lag inversion is therefore present within our energy range. In bin~\emph{d}, $\tau_{12}$, $\tau_{13}$, and $\tau_{14}$ are furthermore all mutually consistent within their uncertainties, indicating that the lag differences across all band pairs become negligible at late times. 

\subsection{GRB~190114C} \label{section_190114C} 

\begin{figure*}
\sidecaption
  \includegraphics[width=12cm]{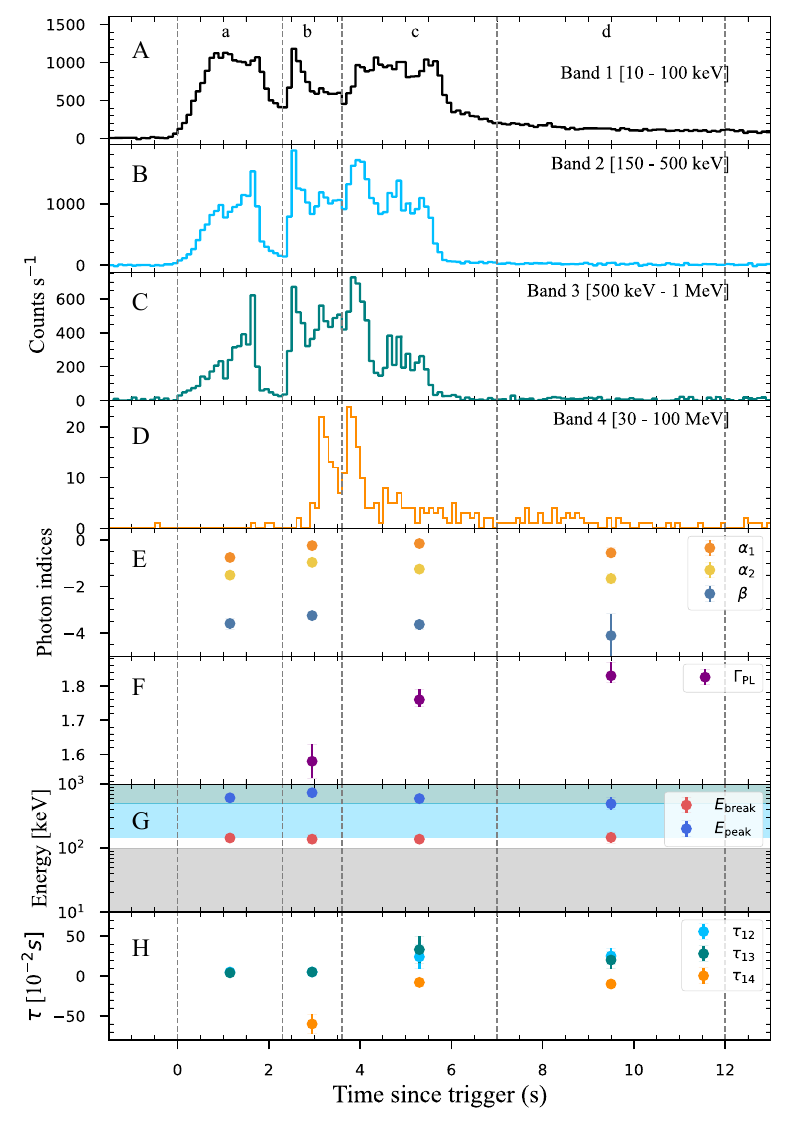}
     \caption{Time evolution of the spectral parameters and time lags obtained from the 2SBPL+PL model (Table~\ref{190114C_results_table}) for the time-resolved spectra of GRB~190114C. Vertical gray dashed lines indicate the time intervals adopted for the spectral analysis (In the time bin \emph{a}, only GBM data were considered). Panels A - D display the count-rate light curves in Bands~1, 2, 3, and~4, respectively, binned at 0.1~s. Panel E shows the photon indices below the break energy, between the break and the peak energies, and above the peak energy (orange, yellow, and blue symbols, respectively). Panel F shows the evolution of the photon index of the additional power-law component (purple symbols). Panel G presents the temporal evolution of $E_\mathrm{peak}$ and $E_\mathrm{break}$ (red and blue symbols, respectively). The light-gray, blue, and green, shaded regions mark the energy ranges corresponding to Bands~1, 2, and~3. Panel H shows the time lags between Band~1 and Bands~2, 3, and~4 (blue, green, and orange symbols, respectively; Table~\ref{time_lags}).}
     \label{parameters_grb190114c}
\end{figure*}

\subsubsection{Time-integrated spectral analysis}

We fitted the time-integrated spectrum of GRB~190114C over the interval 0--12~s from the trigger time with a 2SBPL+PL model. The 2SBPL+PL is strongly preferred over the 2SBPL ($\Delta\mathrm{AIC} \sim 502$; Table~\ref{190114C_results_table}), with the power-law dominating in the LLE energy range at a photon index $\Gamma_\mathrm{PL} \sim 1.85$, consistent with the LAT spectral slope above 100~MeV \citep{2019GCN.23709....1K}. The best fit model is shown in Figure~\ref{spectra_190114C} (upper panel). We discuss the origin of this additional component in Section~\ref{interpretation}.

 \subsubsection{Time-resolved spectral analysis} 

We divided the 0--12~s interval into four bins (vertical dashed lines in Figure~\ref{parameters_grb190114c}), separating the three distinct emission episodes visible in the light curve. In the first bin (0--2.3~s), LLE data were excluded due to insufficient signal, and the spectrum is well described by a 2SBPL only, with no evidence for an additional power-law component.

\begin{figure*}[!ht]
  \centering 
  \includegraphics[width=1.0\textwidth]{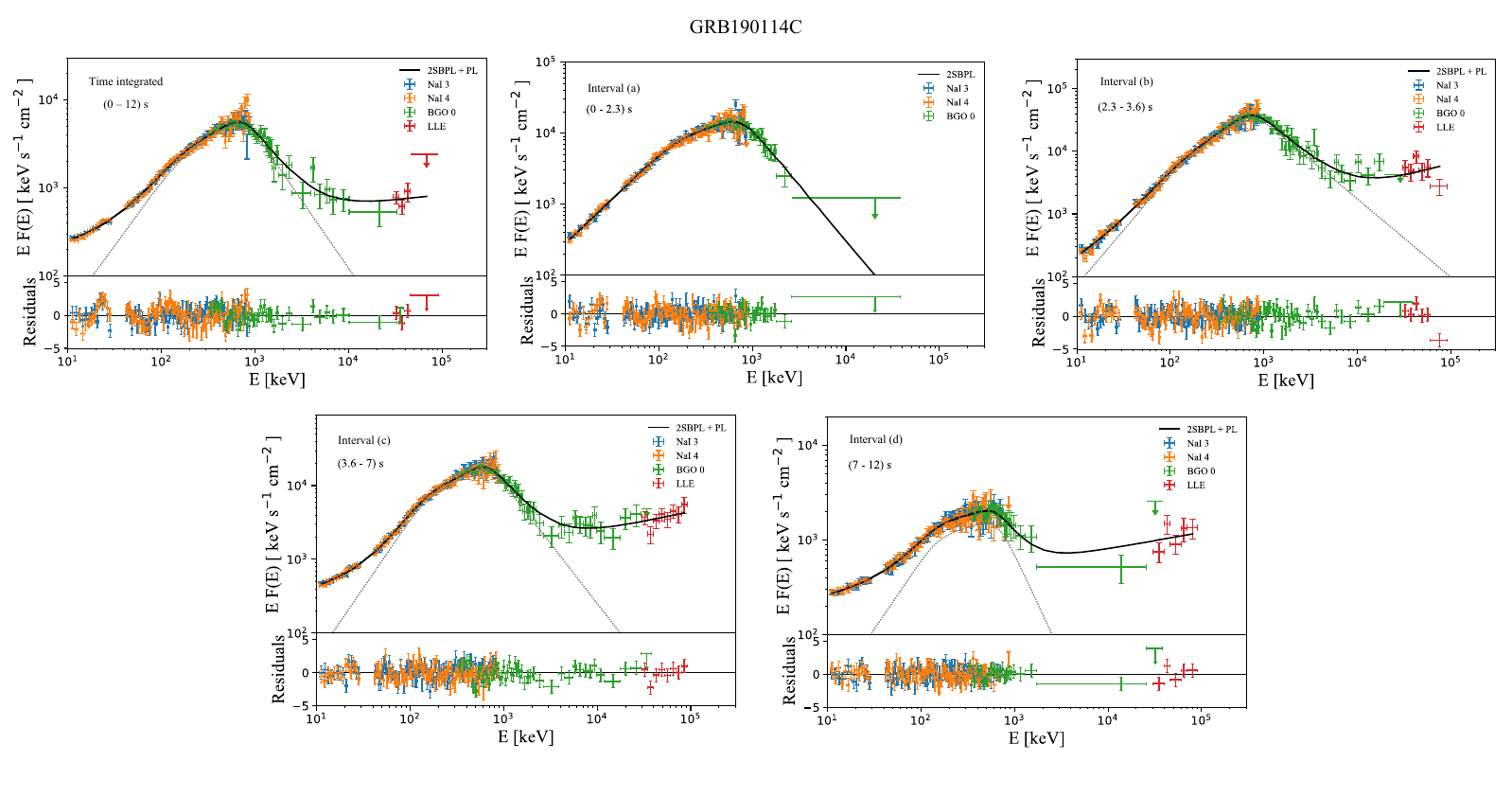}
  \caption{Time-integrated and time-resolved spectra of GRB~190114C in the 10~keV–100~MeV energy range fitted with the 2SBPL+PL model (black solid line). The black-dashed line show the 2SBPL model. Different colors correspond to the individual detectors, as indicated in the legend. In each panel, the lower strip displays the fit residuals. Arrows indicate upper limits.}
  \label{spectra_190114C}
\end{figure*}

From $\sim$2.3~s onward, the 2SBPL+PL model is strongly preferred in all bins ($\Delta\mathrm{AIC} > 100$; see Table~\ref{190114C_results_table}), with the power-law dominating at low energies from $\sim 10-30$ keV, and then from a few tens of MeV across the entire LLE range (Figure~\ref{spectra_190114C}). The value of $\Gamma_\mathrm{PL}$ increases over time (Figure~\ref{parameters_grb190114c}, panel F), with a mean value $\bar{\Gamma}_\mathrm{PL} = -1.75 \pm 0.12$ over 2.3--12~s, consistent with both the LAT measurement above 100~MeV and the result of the time-integrated analysis. Among the 2SBPL parameters, $E_\mathrm{break}$ remains approximately constant around $\sim$141~keV, while $E_\mathrm{peak}$ decreases slowly (Figure~\ref{parameters_grb190114c}, panel G); the photon indices $\alpha_1$, $\alpha_2$, and $\beta$ show no clear temporal trend (Figure~\ref{parameters_grb190114c}, panel E).

\subsubsection{Time-lag analysis}
The evolution of $\tau_{12}$, $\tau_{13}$, and $\tau_{14}$ for GRB~190114C is shown in panel H of Figure~\ref{parameters_grb190114c}, and listed in Table~\ref{time_lags}. No estimate of $\tau_{14}$ was possible in bin \emph{a} due to insufficient LLE statistics.

As for GRB~160625B, $\tau_{12}$ and $\tau_{13}$ are positive, of order $10^{-2}$~s, and mutually consistent; they increase with time, most notably in intervals \emph{b} and \emph{c} ($\sim$3--7~s). In contrast, $\tau_{14}$ is negative in all bins and for the full emission episode, indicating that photons between 30--100~MeV systematically arrive after those at 10--100~keV. Its absolute value decreases sharply with time, from $\sim -0.6$~s in bin \emph{b} to $\sim -0.08$~s in bin \emph{c}, suggesting that the emission process and/or physical conditions evolve significantly between $\sim$3--7~s after the trigger. The burst-integrated value of $\tau_{14}$ is an order of magnitude larger than the per-bin values, likely because the DCF incorrectly associates the bright early pulse visible only in the NaI and BGO light curves ($\sim$2--5~s) with one of the later spiky pulses in the LLE light curve; when these episodes are separated into individual time bins, the inferred $\tau_{14}$ decreases dramatically.

\begin{table*}[!ht]
\renewcommand{\arraystretch}{1.5}
    \centering
    \caption{Best fit parameters for the 2SBPLCUTOFF model for the time-integrated and time-resolved spectra of GRB~160625B.}
    \label{160625B_results_table}
    \begin{adjustbox}{width=\textwidth}
    \begin{tabular}{c c c c c c c c c c c}
        \hline\hline 
         & Time bin & Norm. & $\alpha_1$ & $E_{\text{break}}$ & $\alpha_2$ & $E_{\text{peak}}$ & $\beta$ & $E_\mathrm{cutoff}$ & PG-STAT/dof & $\Delta$AIC\\
         &  [s from trigger time]     & $[\mathrm{photons\,\,keV^{-1}\,\,cm^{-2}\,\,s^{-1}}]$      &            &   [keV]            &            &       [MeV]       &         &  [MeV]       &      \\ 
         \hline
          & time integrated (186 - 205 s) & $9.21^{+0.33}_{-0.34}$  & $-0.77^{+0.01}_{-0.01}$  & $147^{+8}_{-8}$   & $-1.43^{+0.04}_{-0.03}$ & $0.79^{+0.03}_{-0.03}$  & $-2.51^{+0.03}_{-0.02}$  & $62.5^{+12.3}_{-5.0}$  & 689.9/336 & 95.87 \\
          \hdashline
          & a (186 - 189 s)          & $3.98^{+0.45}_{-0.43}$  & $-0.72^{+0.03}_{-0.03}$  & $215^{+34}_{-32}$ & $-1.42^{+0.09}_{-0.09}$ & $2.23^{+0.36}_{-0.28}$ & $-2.35^{+0.11}_{-0.13}$  & $32.4^{+12.6}_{-6.8}$ & 355.8/336 & 31.38 \\
          & b (189 - 192 s)          & $10.5^{+0.7}_{-0.8}$    &  $-0.73^{+0.02}_{-0.02}$ & $148^{+11}_{-12}$ & $-1.54^{+0.06}_{-0.03}$ & $1.15^{+0.06}_{-0.10}$ &  $-2.64^{+0.09}_{-0.03}$ & $ > 111.2$           & 412.8/336 & 6.52 \\
          & c (192 - 199 s)          &  $9.89^{+0.60}_{-0.59}$  & $-0.77^{+0.02}_{-0.02}$  & $133^{+11}_{-11}$ & $-1.49^{+0.05}_{-0.05}$ & $0.61^{+0.03}_{-0.03}$ & $-2.82^{+0.06}_{-0.06}$  & $79.9^{+28.1}_{-26.5}$ & 420.8/336 & 7.20 \\
          & d (199 - 205 s)          & $10.72^{+0.71}_{-0.71}$ & $-0.80^{+0.02}_{-0.02}$  & $155^{+19}_{-19}$ & $-1.38^{+0.06}_{-0.06}$ &  $0.71^{+0.04}_{-0.04}$  & $-2.60^{+0.04}_{-0.04}$  & $> 112.8$           & 496.5/336 & 5.90 \\
         \hline
    \end{tabular}
    \end{adjustbox}
\end{table*}

\begin{table*}[!ht]
\renewcommand{\arraystretch}{1.5}
    \centering
    \caption{Best fit parameters for the 2SBPL+PL model for the time-integrated and time-resolved spectra of GRB~190114C.}
    \label{190114C_results_table}
    \begin{adjustbox}{width=\textwidth}
    \begin{tabular}{c c c c c c c c c c c c}
        \hline\hline 
         & Time bin & Norm. & $\alpha_1$ & $E_{\text{break}}$ & $\alpha_2$ & $E_{\text{peak}}$ & $\beta$ & $\text{Norm}_{\text{PL}}$ & $\Gamma_{\text{PL}}$  & PG-STAT/dof & $\Delta$AIC\\
         &  [s from trigger time]     &  $[\mathrm{photons\,\,keV^{-1}\,\,cm^{-2}\,\,s^{-1}}]$     &            &   [keV]            &            &       [keV]       &         &  $[\mathrm{photons\,\,keV^{-1}\,\,cm^{-2}\,\,s^{-1}}]$    &    &       \\ 
         \hline
          & time integrated (0 - 12 s)  & $1.42^{+0.42}_{-0.51}$ & $-0.54^{+0.10}_{-0.06}$ & $151^{+15}_{-19}$ &  $-1.30^{+0.07}_{-0.05}$ & $656^{+21}_{-35}$ & $-3.54^{+0.23}_{-0.22}$ & $150^{+16}_{-10}$  & $1.85^{+0.03}_{-0.03}$   & 436.3/335 & 502.30\\
          \hdashline
          & a (0 - 2.3 s) - only GBM & $15.8^{+1.2}_{-1.2}$ & $-0.75^{+0.02}_{-0.02}$ & $144^{+11}_{-13}$ & $-1.51^{+0.06}_{-0.05}$ & $612^{+36}_{-30}$ & $-3.59^{+0.20}_{-0.24}$ & - & - & 379.7/330 & - \\
          & b (2.3 - 3.6 s) & $1.27^{+0.78}_{-0.51}$ & $-0.24^{+0.12}_{-0.11}$ & $138^{+28}_{-20}$ & $-0.96^{+0.06}_{-0.07}$ & $735^{+23}_{-22}$ & $-3.25^{+0.12}_{-0.13}$ & $60^{+18}_{-18}$ & $1.58^{+0.05}_{-0.05}$ & 429.0/335 & 150.17 \\ 
          & c (3.6 - 7 s)  & $0.69^{+0.32}_{-0.23}$ & $-0.15^{+0.09}_{-0.08}$ & $138^{+12}_{-10}$ & $-1.25^{+0.06}_{-0.06}$ & $595^{+24}_{-23}$ & $-3.63^{+0.19}_{-0.22}$ & $212^{+21}_{-20}$ & $1.76^{+0.03}_{-0.02}$   & 371.9/335 & 244.61\\
          & d (7 - 12 s)           & $0.76^{+0.89}_{-0.46}$ & $-0.55^{+0.22}_{-0.16}$ & $147^{+29}_{-27}$ & $-1.66^{+0.23}_{-0.19}$ & $494^{+130}_{-93}$ & $-4.11^{+0.95}_{-2.50}$ & $163^{+22}_{-16}$  & $1.83^{+0.04}_{-0.02}$   & 287.7/335 & 174.01 \\
         \hline
    \end{tabular}
    \end{adjustbox}
\end{table*}

\begin{figure*}[!ht]
  \centering
  \includegraphics[width=1.0\textwidth]{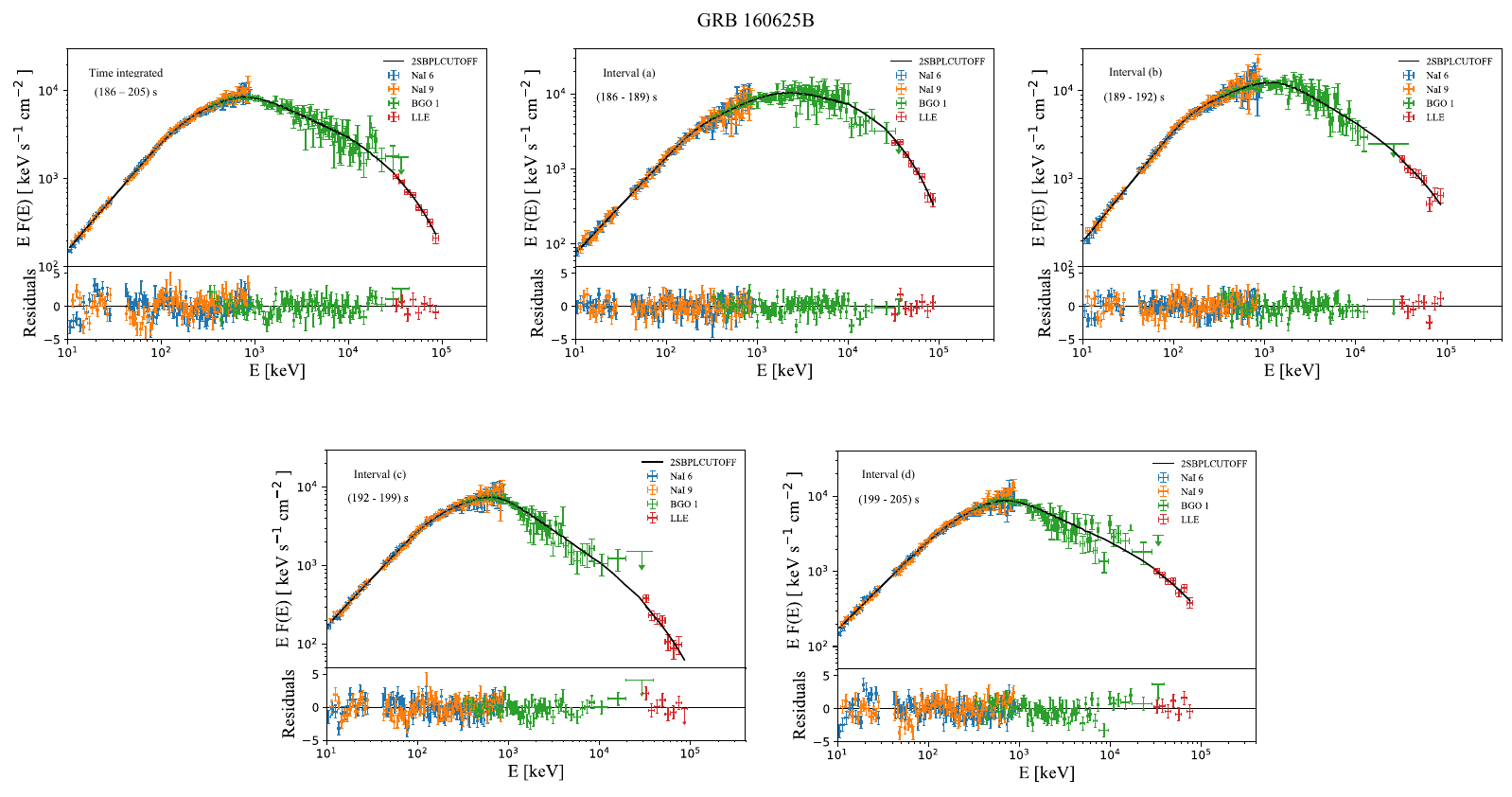}
  \caption{Time-integrated and time-resolved spectra of GRB~160625B in the 10~keV–100~MeV energy range fitted with the 2SBPLCUTOFF model (black solid line). Different colors correspond to the individual detectors, as indicated in the legend. In each panel, the lower strip displays the fit residuals. Arrows indicate lower limits.}
  \label{spectra_grb160625b}
\end{figure*}

\section{Discussion} \label{interpretation}

In this section, we interpret the time-lag and time-resolved spectral results of GRBs~160625B and 190114C (Sections~\ref{section_160625B} and \ref{section_190114C}), and discuss the link between time lags and spectral evolution following the phenomenological scenario initially proposed in M25.

We consider two scenarios, depending on whether one or two spectral components dominate the emission in the considered energy bands.

\textit{(i) Single spectral component.} Time lags arise naturally when a temporally evolving spectrum is observed through finite-width energy bands \citep{mochkovitch2016simple}. For a single component undergoing hard-to-soft evolution, $E_\mathrm{peak}$ shifts progressively to lower energies, causing higher-energy light curves to peak earlier and producing positive lags, a well-established feature of long GRBs \citep{norris2000connection, hakkila2007gamma, hafizi2007time, mochkovitch2016simple}. Crucially, $E_\mathrm{peak}$ does not need cross the band boundaries to produce a measurable lag; continuous spectral softening within a band is sufficient, and this effect can become particularly pronounced when lags are computed between widely separated energy ranges \citep{castignani2014time}.

A high-energy exponential cutoff can modify this picture. If the cutoff remains well above the high-energy band, its effect on time lags is negligible; if it enters or crosses the band, it can suppress the high-energy light curve and produce negative lags. The temporal evolution of the cutoff depends on its physical origin: $\gamma\gamma$ pair-production opacity leads to a cutoff that rises with time as the outflow expands and becomes transparent \citep{ackermann2011detection, granot2008opacity, yassine2017time}, initially suppressing high-energy photons and producing negative lags; conversely, in fast-cooling regime and assuming that the acceleration timescale is proportional to the Larmor time, the observed cutoff energy associated to maximum electron Lorentz factor in the synchrotron component follows the evolution of the bulk Lorentz factor (see e.g. \citealt{piran2010external}), leading to an expected decrease with time, that can be reinforced if the IC cooling becomes important, determining the presence of a positive lag.

\textit{(ii) Two spectral components.} When a second component contributes significantly at high energies, time lags no longer trace the evolution of a single component but instead reflect the relative timing of distinct emission processes. The most common scenario is a delayed high-energy component, which may be of prompt origin, or associated with the afterglow: if low-energy bands are dominated by the prompt emission only, while the high-energy band is dominated by a component that emerges later, the high-energy light curve peaks later, producing negative lags \citep{ackermann2013multiwavelength, ajello2019decade, 2019ICRC...36..555B}. The opposite configuration of an early-peaking high-energy component is physically less natural, as high-energy components from external shocks or opacity-regulated processes typically emerge progressively later than the main component. 

In summary, positive lags most naturally indicate a single dominant component undergoing hard-to-soft spectral evolution, while negative lags are intrinsically more ambiguous: they may arise from an evolving $\gamma\gamma$ cutoff within a single component, or from the delayed onset of a second high-energy component. Disentangling these scenarios requires a joint spectral and temporal analysis.

\subsection{GRB~160625B: hard-to-soft evolution of a single component} \label{subsec:interp_160625b}

The time-resolved spectral analysis of GRB~160625B shows that the emission is dominated throughout by a single 2SBPLCUTOFF component. $E_\mathrm{peak}$ and $E_\mathrm{break}$ both decrease with time, exhibiting a clear hard-to-soft evolution. The time lags are positive across the full 10~keV--100~MeV range: $\tau_{12}$ and $\tau_{13}$ are small and mutually consistent ($\sim 10^{-2}$~s), while $\tau_{14}$ 
decreases slowly with time. This behavior is consistent with the hard-to-soft evolution of the prompt emission: the wider energy separation between the LLE and GBM bands amplifies the lag, and its gradual decrease reflects the slowing spectral evolution as $E_\mathrm{peak}$ shifts to lower energies at later times.

However, we note that in bins~\emph{c} and~\emph{d}, $\tau_{14}$ decreases to values smaller than $\tau_{12}$ and $\tau_{13}$ (Table~\ref{time_lags}). To assess whether this implies a lag inversion between Band~4 and Bands~2--3, we explicitly computed the lags $\tau_{24}$ and $\tau_{34}$ between Band 2/3 and Band 4 in these two intervals, respectively. While the values of $\tau_{24}$ and $\tau_{34}$ computed in intervals \emph{c} and \emph{d} are found to be negative, they are all compatible with zero within their uncertainties, so that no statistically significant lag inversion is detected between energy ranges involved in this work. We caution, however, that these results are affected by the limited photon statistics available at high energies in the LLE band, and that higher-quality data would be needed to draw more definitive 
conclusions on this point. We also note that a transition from positive to negative lags in GRB~160625B has been previously reported by \citet{wei2017new}, though using energy bands extending up to LAT energies ($>$100~MeV), i.e., beyond the LLE range. The absence of a statistically significant inversion within our energy range is therefore consistent with both the literature and the single hard-to-soft component scenario we propose for GRB~160625B. 

A natural explanation for this trend, within the single-component 
scenario, is suggested by the analytical model of 
\citet{mochkovitch2016simple}, who propose that the sensitivity of 
a given lag to each spectral parameter depends on the position of 
the energy bands relative to $E_\mathrm{peak}$: when both bands 
lie below $E_\mathrm{peak}$, the lag is primarily controlled by 
the evolution of $\alpha_1$ and $\alpha_2$, whereas when one band 
lies above $E_\mathrm{peak}$, the lag becomes sensitive to the 
evolution of $\beta$. In bins~\emph{c} and~\emph{d}, Band~4 
(30--100~MeV) lies above $E_\mathrm{peak}$ ($\sim$0.6--0.7~MeV), 
while Bands~1--3 remain below it, so that $\tau_{12}$ and 
$\tau_{13}$ are mainly governed by the evolution of 
$\alpha_1$ and $\alpha_2$, while $\tau_{14}$ should be more 
sensitive to the evolution of $\beta$. According to this model, 
an increase in $\beta$ could therefore reduce $\tau_{14}$ relative 
to $\tau_{12}$ and $\tau_{13}$. Indeed, from interval~\emph{c} 
to~\emph{d}, $\beta$ slightly increases 
(Table~\ref{160625B_results_table}).

The high-energy cutoff, detected at $\sim$100~MeV across all 
intervals, does not significantly affect the measured lags, since 
the LLE emission of GRB~160625B is concentrated between 30--60~MeV, 
well below the cutoff. Whether the cutoff is arising due to 
$\gamma\gamma$ pair-production opacity or, in the case the 
high-energy prompt emission is due to IC scatterings, from the 
intrinsic spectral shape of this additional component, is 
investigated in Appendix~\ref{computing_gamma}, where we estimate 
the bulk Lorentz factor $\Gamma$ of the emitting region for these 
two assumptions. If the cutoff arises from photon--photon pair 
production, we find $\Gamma_{\gamma\gamma} \sim 120$ in the first 
interval, increasing to $\sim$230--260 in subsequent intervals, 
with a time-integrated value of $\Gamma_{\gamma\gamma} \simeq 230$ 
(see Table~\ref{lorentz_factor}). The emission radius $R_\mathrm{MeV} 
\sim 10^{14}$~cm remains approximately constant in time and satisfies 
$R_\mathrm{MeV} \gtrsim R_\mathrm{ph} \sim 10^{13}$~cm in all 
intervals (Fig~\ref{lorentfactor_160625b}, panel C), confirming that 
the prompt emission is produced in an optically thin region above 
the photosphere, as expected in internal shock or some magnetic 
reconnection scenarios. However, sub-photospheric models cannot be 
excluded on the basis of these constraints alone 
\citep{ryde2010identification, ahlgren2019testing}. If the cutoff 
reflects the intrinsic curvature of the IC spectrum, only lower 
limits on $\Gamma$ can be derived; these closely track the 
$\Gamma_{\gamma\gamma}(E_\mathrm{cutoff})$ values while remaining 
systematically larger (Fig~\ref{lorentfactor_160625b}, panel B).

The case of GRB~160625B is consistent with a scenario where the observed positive time lags are linked to the hard-to-soft spectral 
evolution of a single dominant emission component. The time-resolved 
spectral analysis identifies this component with the prompt emission, 
whose spectral properties and transparency condition suggest the 
emission originates in an optically thin region above the photosphere.

\subsection{GRB~190114C: delayed emergence of a high-energy component} \label{subsec:interp_190114c}

In GRB~190114C, an additional power-law spectral component emerges at $\sim$2.5~s after the trigger and dominates the LLE energy range and the soft X-ray energy range thereafter. Correspondingly, $\tau_{14}$ is negative in all time bins, indicating that 30--100~MeV photons arrive systematically later than those at 10--100~keV, while $\tau_{12}$ and $\tau_{13}$ remain positive and of order $10^{-2}$~s. The negative $\tau_{14}$ is thus a direct consequence of the delayed high-energy component, rather than the spectral evolution of the prompt.

\begin{figure}[!ht]
  \centering
  \includegraphics[width=0.5\textwidth]{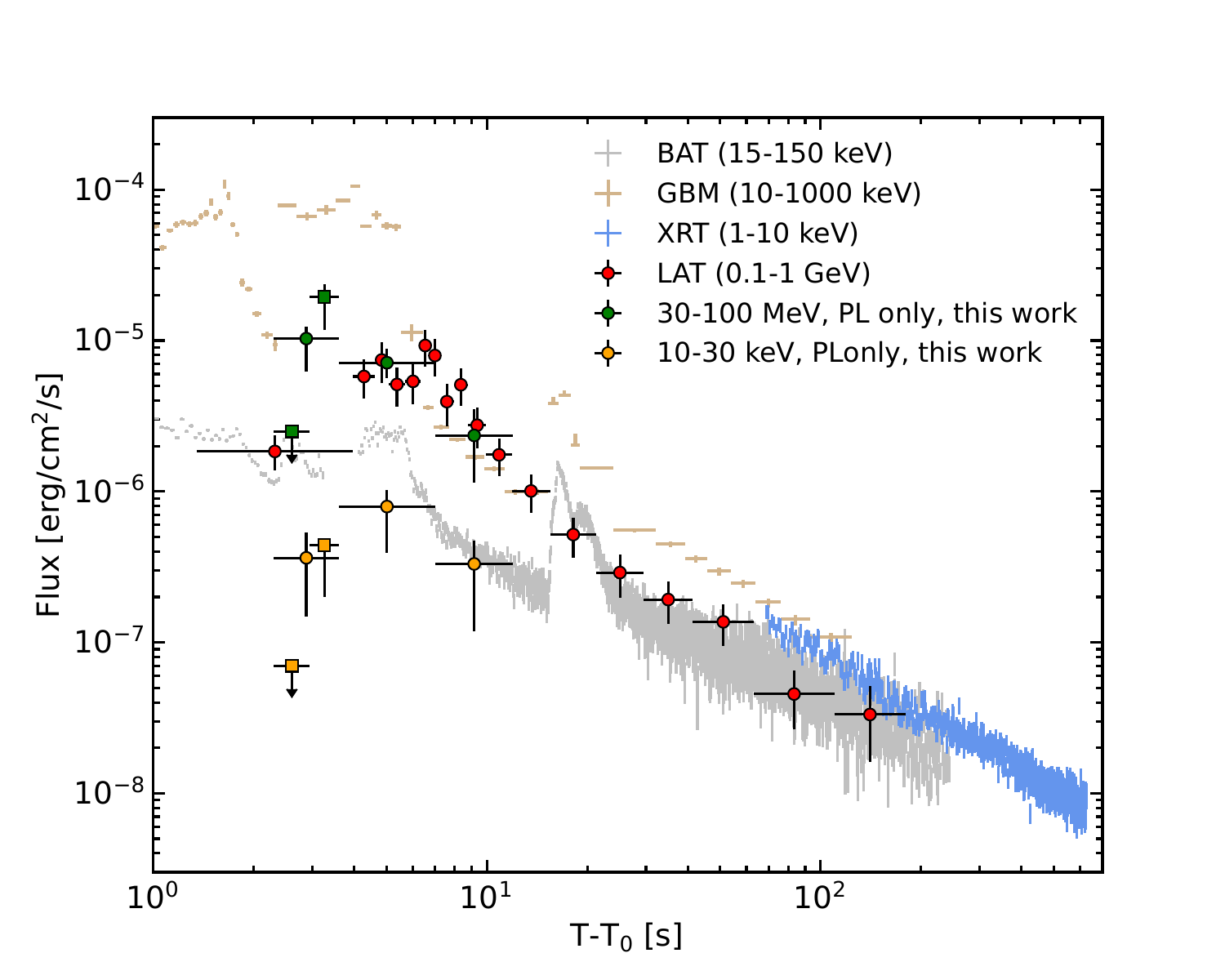}
  \caption{Lightcurves of GRB~190114C across different energy bands: \emph{Swift}-XRT (light blue crosses), \emph{Swift}-BAT (gray crosses), GBM (gold crosses), \emph{Fermi}-LAT data (red filled circles). The flux of the powerlaw component is shown with green (30-100\,MeV) and yellow (10-30\,keV) circles. The results obtained after dividing the first temporal bin into two sub-bins are shown with filled squares. Points represented by arrows indicate upper limits.}
  \label{190114C_lightcurves}
\end{figure}

A possible explanation for the extra component dominating both at low and high energies is a contribution from an emerging afterglow emission. To test the consistency of this interpretation, we compare the flux of this spectral component with the flux of early time afterglow emission. 
Observations from BAT and GBM showed the presence of a powerlaw decaying flux (starting from $\sim
\,25\,$s), interpreted as synchrotron afterglow radiation, whose smooth temporal decay forms an underlying continuum beneath the variability of the remaining prompt emission peaks (see gray and gold crosses in Figure~\ref{190114C_lightcurves}), interpreted as synchrotron afterglow radiation. The afterglow is also visible with XRT (although observations start at later times, $\sim$\,70\,s) and LAT (0.1-1\,GeV), although a hint of  variability at early times questions the presence of a contribution from internal radiation up to $\sim$\,10 seconds, after which the GeV light curve is consistent with a smooth, powerlaw temporal decay of external origin.

\begin{table}[!ht]
\renewcommand{\arraystretch}{1.3}
    \centering
    \caption{Time lags of GRB~160625B and ~190114C, computed across 
    their whole emission episode and in the same time-bins adopted 
    for the spectral analysis, corresponding to separate pulses.}
    \label{time_lags}
    \begin{adjustbox}{width=\columnwidth}
    \begin{tabular}{l c c c c}
        \hline\hline 
        Time interval & $\tau_{12}$ $(10^{-2}$~s$)$ & $\tau_{13}$ $(10^{-2}$~s$)$ & $\tau_{14}$ $(10^{-2}$~s$)$ \\
        {[s from trigger time]} & & & \\
        \hline
        \multicolumn{4}{l}{GRB~160625B} \\
        \hline
        time integrated (186--205 s) & $8 \pm 5$ & $13 \pm 4$ & $66 \pm 12$ \\
        \hdashline
        a (186--189 s)   & $5 \pm 2$ & $6 \pm 2$  & $25 \pm 8$  \\
        b (189--192 s)   & $7 \pm 3$ & $6 \pm 3$  & $13 \pm 5$  \\
        c (192--199 s)   & $9 \pm 3$ & $8 \pm 3$  & $4 \pm 1$   \\
        d (199--205 s)   & $9 \pm 4$ & $9 \pm 5$  & $6 \pm 2$   \\
        \hline
        \multicolumn{4}{l}{GRB~190114C} \\
        \hline
        time integrated (0--12 s) & $11 \pm 4$ & $13 \pm 7$ & $-229 \pm 46$ \\
        \hdashline
        a (0--2.3 s)     & $5 \pm 2$  & $4 \pm 2$  & --          \\
        b (2.3--3.6 s)   & $5 \pm 2$  & $5 \pm 3$  & $-60 \pm 12$ \\
        c (3.6--7 s)     & $24 \pm 15$& $33 \pm 17$& $-8 \pm 5$  \\
        d (7--12 s)      & $25 \pm 10$& $20 \pm 11$& $-10 \pm 3$ \\
        \hline
    \end{tabular}
    \end{adjustbox}
\end{table}

To assess the consistency with afterglow emission of the extra powerlaw component present during the prompt emission, we estimate the flux of this extra component and compare it with early time observations of afterglow radiation.
We calculate the flux of the PL component in two specific energy ranges where the powerlaw dominates over the 2SBPL component (see Figure \ref{spectra_190114C}): 10--30 keV and 30--100 MeV. Fluxes are calculated for all time intervals b--d defined in our time-resolved spectral analysis. Notably, we further divided bin b in two sub intervals, to better resolve the flux evolution within the first $\sim 3\,$s of the emission. We find that in the first sub-bin, the additional PL component is not statistically required (i.e., the data are well-described by a single 2SBPL model). In this case, we derive a $3\sigma$ upper limit to the powerlaw flux. This is computed by fixing the parameters obtained by the fit of the 2SBPL component  and adding a powerlaw with fixed spectral index (see below) and increasing  normalization until the fit worsened by a $3\sigma$ significance level compared to the pure 2SBPL model. The resulting normalization is used to calculate the flux upper limit in the two energy bands (10--30 keV and 30--100 MeV). For the spectral index of the powerlaw we considered the softest and the hardest values obtained  in time resolved analysis ($\Gamma_{\mathrm{PL}} = 1.83$ and 1.39), and select the  most conservative result on the flux upper limit derived.
Figure~\ref{190114C_lightcurves} shows lightcurves across different energy bands, including low-energy data from \emph{Swift}-XRT (light blue crosses) and \emph{Swift}-BAT (gray crosses), GBM (gold crosses) alongside high-energy \emph{Fermi}-LAT data (red filled circles). The flux of the powerlaw component is shown with green (30-100\,MeV) and yellow (10-30\,keV) circles. The results obtained after dividing the first temporal bin into two sub-bins are shown with filled squares.

Our multi-wavelength comparison shows a good compatibility between the powerlaw flux and the afterglow emission observed by \emph{Swift}, GBM, and \emph{Fermi}-LAT. However, the rise of the powerlaw flux is very steep (steeper than $t^{3}$), in tension with an interpretation as onset of the afterglow. The physical nature of this component at early times remains uncertain and might require invoking a contribution from an internal high-energy component. Consequently, two physical interpretations remain viable. The additional component may be associated with the onset of the early afterglow from external shocks \citep{ravasio2019grb, ronchi2020rise, ajello2020fermi}, in which case the observed delay reflects the deceleration timescale of the outflow and the temporal evolution of $\tau_{14}$ may constrain the afterglow onset. Alternatively, the delayed component could arise from synchrotron self-Compton (SSC) emission within the same internal dissipation region as the prompt, with the delay driven by the evolving SSC efficiency \citep{daigne2003physics, bovsnjak2009prompt, bovsnjak2014spectral} or the gradual reduction of early-time $\gamma\gamma$ opacity \citep{lithwick2001lower, hascoet2012fermi, yassine2017time}.

Regardless of the specific mechanism and on the origin of the additional power-law component, the case of GRB~190114C illustrates the scenario in which negative time lags arise due to the presence of different emission components dominating in different energy ranges. These results highlights the need for a joint spectral and temporal analyses to reliably interpret the origin of such component.

\subsection{The case of a single spectral component with evolving HE absorption } \label{subsec:bias}

A third scenario, not applicable to any of the two GRBs analyzed in this work, ascribes negative lags to a single component whose high-energy cutoff evolves due to $\gamma\gamma$ opacity. In this scenario a second spectral component is not required. In this case, high-energy photons are initially suppressed by the high photon density of the compact emitting region, and escape only as the outflow expands and becomes transparent, naturally producing negative lags.

Such case is not found among the brightest GRBs in the sample of M25, i.e. the ones for which a spectral analysis can be performed, likely reflecting an observational bias. Measurable negative lags from a time-dependent $\gamma\gamma$ cutoff require high optical depths at early times, implying comparatively low bulk Lorentz factors \citep{lithwick2001lower, hascoet2012fermi}. Since $\Gamma$ correlates with the isotropic-equivalent luminosity \citep{ghirlanda2010onset, ghirlanda2018bulk}, low-$\Gamma$ GRBs are intrinsically fainter and may be not feasible for a spectral analysis, or even fall below the LLE detection threshold, introducing a natural bias against their inclusion in LLE-selected samples. A more quantitative assessment of the conditions under which evolving $\gamma\gamma$ opacity can generate observable negative lags, and of the associated selection effects, will be addressed in a dedicated theoretical study to be presented in a future work.

\section{Conclusions} \label{discussion} 

We have investigated the connection between time lags and spectral 
evolution in GRBs by combining joint GBM+LLE time-resolved spectral 
analyses with time-lag measurements for GRBs~160625B and 190114C, 
two bright \emph{Fermi} events with high-quality broadband data from 
keV to tens of MeV and multiple distinct emission episodes. Building 
on the phenomenological framework of M25, we find that time lags 
measured between light curves in different energy bands provide a 
complementary, largely model-independent diagnostic of GRB spectral 
evolution. Unlike spectral fitting, which requires sufficient photon 
statistics and broad energy coverage, time lags can be robustly 
estimated even when detailed spectral modeling is not feasible, 
making them a valuable proxy for spectral evolution.

For GRB~160625B, the emission is dominated throughout by a single non-thermal component (2SBPLCUTOFF), with $E_\mathrm{break}$ and $E_\mathrm{peak}$ shifting to lower energies over time in a clear hard-to-soft evolution. All measured lags are positive across the full 10~keV--100~MeV range: $\tau_{12}$ and $\tau_{13}$ are small and comparable throughout all intervals, while $\tau_{14}$ is larger in the early intervals but decreases to values comparable or smaller than $\tau_{12}$ and $\tau_{13}$ in bins~\emph{c} and~\emph{d}. According to the analytical model of lags from \citet{mochkovitch2016simple}, since $E_\mathrm{peak}$ always lies below the LLE band, $\tau_{14}$ is sensitive to the evolution of $\beta$ throughout the burst, while $\tau_{12}$ and $\tau_{13}$ are primarily governed by the evolution of $\alpha_1$ and $\alpha_2$. The decrease of $\tau_{14}$ in bins~\emph{c} and~\emph{d} is therefore consistent with $\beta$ increasing in those intervals, since this primarily affects $\tau_{14}$ without significantly 
changing $\tau_{12}$ and $\tau_{13}$. The lags $\tau_{24}$ and $\tau_{34}$ between Bands 2/3 and Band 4, computed in bins~\emph{c} and~\emph{d}, are all compatible with zero within $1\sigma$, confirming that no statistically significant lag inversion is present at these energies. The high-energy cutoff near $\sim$100~MeV does not affect the lags, as the LLE emission is concentrated between 30--60~MeV and the transparency condition $R_\mathrm{MeV} \gtrsim R_\mathrm{ph}$ is satisfied in all intervals (Table~\ref{lorentfactor_160625b}). GRB~160625B therefore provides a case consistent with a scenario in which positive time lags are linked to the hard-to-soft spectral evolution of a single dominant prompt emission component.

For GRB~190114C, the low-energy emission is similarly described by a 2SBPL, but an additional power-law component emerges at $\sim$2.5~s and dominates the LLE range, so that time-integrated and time-resolved spectra require a 2SBPL+PL model. The lags $\tau_{12}$ and $\tau_{13}$ remain small and positive, suggesting that the keV--MeV emission is dominated by a single component, while $\tau_{14}$ is negative in all time bins, indicating that 30--100~MeV photons arrive systematically later. This behavior naturally arises when distinct spectral components dominate different energy bands. The delayed high-energy component may originate from external shocks (early afterglow) or from internal dissipation such as SSC emission, with early-time $\gamma\gamma$ opacity potentially contributing to the delay. Our multi-wavelength comparison confirms that the flux of this isolated powerlaw component is broadly consistent with the afterglow emission observed by \emph{Swift} and \emph{Fermi}-LAT, although early-time LAT variability suggests a more complex onset. Negative lags are therefore intrinsically more ambiguous than positive ones, and their interpretation requires detailed spectral analysis combined with physical modeling. GRB~190114C nonetheless provides a clear example in which negative lags reflect the delayed emergence of a distinct high-energy spectral component.

We also find that none of the bright LLE-detected GRBs with negative $\tau_{14}$ in M25 can be modeled by a single spectral component alone. This is consistent with a natural observational bias: producing measurable negative lags in a single-component scenario requires a time-dependent $\gamma\gamma$ cutoff, implying low bulk Lorentz factors and thus intrinsically fainter bursts that are less likely to be detected or analyzed in the LLE range. A quantitative assessment of this scenario and its associated selection effects will be addressed in a future work.

Notably, 40\% of the 70 GRBs in the M25 sample display a significantly positive $\tau_{14}$, suggesting that this lag behavior is common across the GRB population. Whether it universally reflects the same single-component hard-to-soft evolution as in GRB~160625B, however, cannot be established without time-resolved spectral analysis of each individual burst, a task that is currently limited to the brightest LLE-detected events. Similarly, 37\% of the M25 sample exhibits a significantly negative $\tau_{14}$, confirming that the delayed onset of a high-energy component is a recurring phenomenon. The physical origin of this behavior in each individual case, be it early afterglow emission, SSC radiation, or another process, remains to be determined through dedicated broadband spectral studies of the relevant sub-population.

In conclusion, our results support a unified phenomenological picture in which positive time lags trace the hard-to-soft spectral evolution of a single dominant emission component, while negative lags most naturally signal the delayed emergence of an additional high-energy component. Time lags alone are insufficient to unambiguously identify the origin of the emission, but combined with time-resolved spectral analysis they constitute a powerful diagnostic of the physical processes at work. Future progress will benefit from larger GRB samples with high-quality broadband coverage, improved high-energy sensitivity, and joint temporal--spectral modeling with physically motivated emission models.

\begin{acknowledgements}
C.M. is grateful to the Institut d'astrophysique de Paris (IAP) for their warm hospitality during the course of this project. LN and GG acknowledge  the European Union-Next Generation EU, PRIN 2022 RFF M4C21.1 (202298J7KT - PEACE). This work has been supported by the project GRB PrOmpt Emission Modular Simulator (POEMS) financed by INAF Grant 1.05.23.06.04.
\end{acknowledgements}

\bibliographystyle{aa} 
\bibliography{aa60232-26} 

\begin{appendix}

\section{Estimation of the Lorentz factor for GRB~160625B} \label{computing_gamma} 

\begin{table*}[!ht]
    \centering
    \caption{Spectral parameters and results of the Lorentz factor computation, for all the time-intervals of GRB~160625B analyzed.}
    \label{lorentz_factor}
     \begin{adjustbox}{width=\textwidth}
    \begin{tabular}{l l c c c c c c c}
         \hline\hline 
         & Time interval (s from trigger time)   & time integrated (186 - 205 s) & a (186 - 189 s) & b (189 - 192 s)  & c (192 - 199 s) & d (199 - 205 s) \\
         \hline
         & $E_\mathrm{cutoff}$ (MeV)                                                     & $62.5^{+12.3}_{-5.0}$ & $32.4^{+12.6}_{-6.8}$ & $> 111.2$ & $79.9^{+28.1}_{-26.5}$ & $> 112.8$\\
         & $E_*$ (MeV)                                                                & 51 & 44  & 56 & 67 & 50\\
         & $F(E_*)\,\,(10^{-4}\, \text{cm}^{-2}\, \text{MeV}^{-1})$                   & $4.58 \pm 0.86$ & $2.20 \pm 0.75$ & $6.35 \pm 0.43$ & $4.38 \pm 0.54$ & $6.78 \pm 0.95$\\
         & $s$                                                                        & $-2.51 \pm 0.03$ & $-3.72 \pm 0.05$ & $-2.64 \pm 0.06$ &  $-2.82 \pm 0.06 $ &  $-2.60 \pm 0.04$\\
         & $\phi (s)$                                                                 & $0.453 \pm 0.004$ & $0.437 \pm 0.005$ & $0.450 \pm 0.004$ & $0.445 \pm 0.003$ & $0.451 \pm 0.003$\\
         & Emission radius $R_{\text{MeV}} \,(10^{14}\,\text{cm})$                    & $1.29 \pm 0.53$ & $1.36 \pm 0.45$ & $ >1.73 $& $1.32 \pm 0.35$ & $ > 1.69 $\\
         & Lorentz factor $\Gamma_{\gamma\gamma} (E_\mathrm{cutoff})$                 & $227.14 \pm 15.43$ & $119.43 \pm 12.34$ & $> 256.87$ & $229.95 \pm 11.88$ & $ >252.95$\\
         & Luminosity $(10^{53} \text{erg} \, \text{s}^{-1})$                         & $3.03 \pm 0.99$ & $2.36 \pm 1.01$ & $4.11 \pm 1.56$ & $2.87 \pm 0.76$ & $5.09 \pm 1.43$\\
         & Photospheric radius $R_{\text{ph}} \,(10^{14}\,\text{cm})$                 & $0.10 \pm 0.01$ &  $0.51 \pm 0.19$ & $< 0.09$ & $0.09 \pm 0.02$ & $< 0.11$ \\
         & $E_{\text{max}}$ (MeV)                                                     & 99.83 &  99.46 & 99.83 & 99.67 & 99.47\\
         & Lower limit on the Lorentz factor $\Gamma_{\gamma\gamma} (E_{\text{max}})$ & $251.22 \pm 25.03$ & $163.31 \pm 23.67$ & $263.19 \pm 19.87$ & $243.04 \pm 18.65$ & $260.11 \pm 19.22$\\
         \hline           
    \end{tabular}
    \end{adjustbox}
    \tablefoot{The values of $\Gamma_{\gamma\gamma}(E_\mathrm{cutoff})$ and $R_\mathrm{MeV}$, and $\Gamma_{\gamma\gamma}(E_\mathrm{max})$ refers to the two alternative interpretations described in this section: either the high-energy cutoff at $E_\mathrm{cutoff}$ is due by gamma-gamma opacity to pair production, which allows us to provide a direct measurement of the Lorentz factor, or the cutoff represents an intrinsic spectral break, in which case only a lower limit on the Lorentz factor can be derived considering the maximum energy of the observed photons, $E_\mathrm{max}$. }
\end{table*}

\begin{figure*}
  \centering
  \includegraphics[width=12cm]{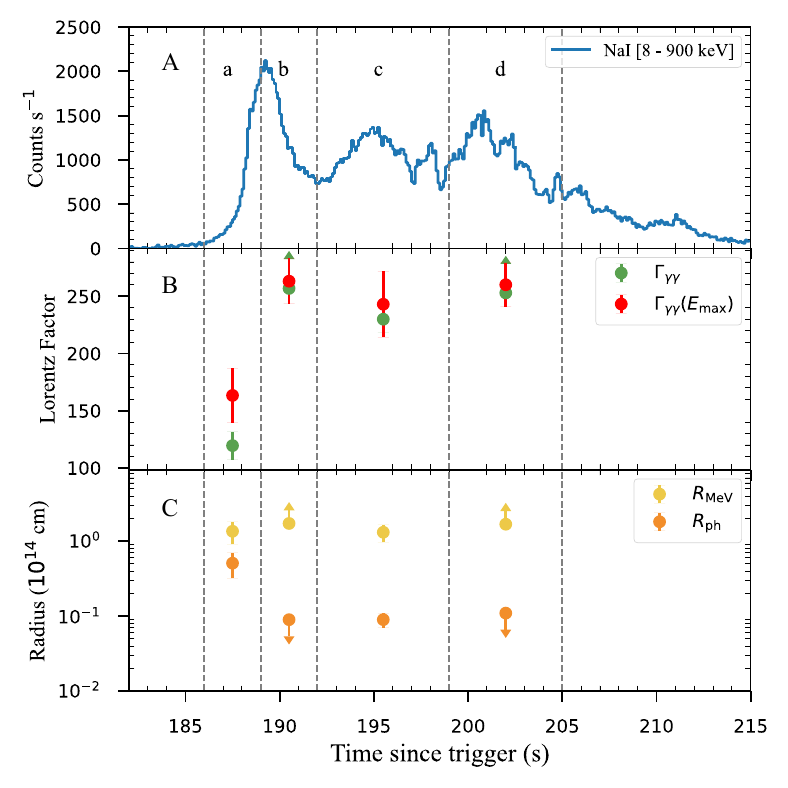}
     \caption{Time evolution of the Lorentz factor and characteristic radii, for the time-resolved spectra of GRB~160625B. Vertical gray dashed lines indicate the time intervals adopted for the spectral analysis. Panel A displays the count-rate light curve in the full NaI energy range, binned at 0.1~s. Panel B shows the Lorentz factors measured assuming that the high-energy cutoff at $E_\mathrm{cutoff}$ is either due to gamma-gamma opacity to pair production, or a natural break of the emission mechanisms (green and red symbols, respectively). Panel C presents the evolution of the radius at which the MeV photons are produced, and the photospheric radius (yellow and orange symbols, respectively). Arrows indicate upper or lower limits.}
     \label{lorentfactor_160625b}
\end{figure*}

The high-energy exponential cutoff observed in the time-integrated and time-resolved spectra of GRB~160625B (see Table~\ref{160625B_results_table} and Figure~\ref{spectra_grb160625b})  can be interpreted in two ways: if the cutoff arises from photon–photon pair production ($\gamma\gamma \rightarrow \mathrm{e^{+}e^{-}}$), the bulk Lorentz factor $\Gamma$ of the emitting region can be directly estimated \citep{lithwick2001lower, pe2007new, hascoet2012fermi, yassine2017time, ravasio2024insights}. In the hypothesis that the high-energy prompt-emission spectrum is associated with an IC origin, the spectral break can reflect an intrinsic spectral curvature from IC scattering in the Klein–Nishina regime (hereafter natural break), only a lower limit on $\Gamma$ can be derived. In this section, we estimate $\Gamma$ for both scenarios following the methodology presented in \citet{yassine2017time}, assuming throughout that the prompt emission originates above the photosphere in an optically thin regime, and verifying this assumption a posteriori.

\subsubsection*{Case (i): Cutoff due to $\gamma$-ray opacity to pair creation}

In the optically thin regime, the radius at which MeV photons are produced is given by 

\begin{equation} \label{radius_mev}
  R_\mathrm{MeV} \simeq 2c\Gamma^2\frac{t_\mathrm{v}}{1+z}
\end{equation}

valid for internal shocks \citep{daigne1998gamma, rees2005dissipative} and some magnetic reconnection models (e.g., ICMART, \citealt{zhang2010internal}). If the high-energy cutoff is due to photon opacity to pair creation, the bulk Lorentz factor $\Gamma_{\gamma\gamma}$ is then estimated from \citet{hascoet2012fermi}:

\begin{equation} \label{gamma_formula}
\begin{split}
        \Gamma_{\gamma\gamma} = \frac{K \Phi(s)}{\left[\frac{1}{2}\left(1 + \frac{R_{\mathrm{GeV}}}{R_\mathrm{MeV}}\right) \left(\frac{R_{\mathrm{GeV}}}{R_\mathrm{MeV}}\right)\right]^{1/2}}\,\,\left(1 + z\right)^{-(1+s)/(1-s)} \times\\
        \times\,\left\{\sigma_\mathrm{T}\left[\frac{D_\mathrm{L}(z)}{c\,t_{\mathrm{v}}}\right]^2 E_* F(E_*)\right\}^{1/2(1-s)}\left[\frac{E_*E_\mathrm{cut}}{(m_ec^2)^2}\right]^{(s+1)/2(s-1)}.
\end{split}
\end{equation}

where $\sigma_\mathrm{T}$ is the Thomson cross section, $D_\mathrm{L} = 3.17 \times 10^{28}$\,cm is the luminosity distance at $z = 1.406$ \citep{2016GCN.19600....1X}. The constant $K \simeq 0.4$–$0.5$ is calibrated via numerical calculations of the $\gamma\gamma$ opacity process \citep{hascoet2012fermi}; we let $K$ vary across this range and adopt the corresponding average $\Gamma_{\gamma\gamma}$ as our result. For simplicity, we allow the variability timescale $t_\mathrm{v}$ to vary within the range 0.1–1 s, which is consistent with what is obtained for long \emph{Fermi} GRBs \citep{golkhou2015energy}. We also assume that MeV and GeV photons are produced in the same region ($R_\mathrm{MeV} \simeq R_\mathrm{GeV}$).

The quantities needed to compute $\Gamma_{\gamma\gamma}$ are listed in Table~\ref{lorentz_factor}: $E_\mathrm{cutoff}$ is the cutoff obtained from the 2SBPLCUTOFF fit; $E_*$, the characteristic seed photon energy interacting with cutoff-energy photons \citep[Eq.~8 in][]{yassine2017time}; $F(E_*)$, the photon fluence at $E_*$ over $t_\mathrm{v}$; $s$, the local photon index around $E_*$; and $\Phi(s)$, a function of $s$ alone \citep[Eqs.~9--10 in][]{yassine2017time}. Since the seed spectrum is locally a power law, $\Gamma_{\gamma\gamma}$ is insensitive to the specific choice of $E_*$. The index $s$ was estimated via Monte Carlo sampling of the covariance matrix from the 2SBPLCUTOFF fits. Similarly, the corresponding values of $F(E_*)$ were obtained directly from the spectral fits.

The resulting $\Gamma_{\gamma\gamma}(E_\mathrm{cutoff})$ and $R_\mathrm{MeV}$ values are shown in Figure~\ref{lorentfactor_160625b} (green and yellow symbols, respectively) and listed in Table~\ref{lorentz_factor}. The bulk Lorentz factor increases from $\sim$120 in interval \emph{a} to $\sim$230--260 in subsequent intervals; lower limits apply in intervals \emph{b} and \emph{d} where $E_\mathrm{cutoff}$ is unconstrained. The time-integrated value $\Gamma_{\gamma\gamma} \simeq 230$ is consistent with the time-resolved average. The emission radius $R_\mathrm{MeV} \sim 10^{14}$~cm is approximately constant across intervals, with lower limits in intervals \emph{b} and \emph{d}.

To verify the self-consistency of the optically thin assumption, we compare $\Gamma_{\gamma\gamma}$ with the minimum Lorentz factor required for transparency to Thomson scattering, using the photospheric radius \citep{beloborodov2013regulation}:

\begin{equation}
    R_\mathrm{ph} \simeq \frac{\sigma_\mathrm{T} \left(1 + f_\pm\right)\dot{E}}{8 \pi c^3m_\mathrm{p} \bar{\Gamma}^3(1+\sigma)}
\end{equation}

where $\bar{\Gamma} \simeq \frac{1+k}{2}\,\Gamma_{\gamma\gamma}$ is the mean outflow Lorentz factor (with $k \sim 2$–$5$ being the contrast factor; \citealt{hascoet2012fermi}); $f_\pm$ is the pair-to-electron ratio, which vanishes when the pair-creation optical depth is below unity at $R_\mathrm{MeV}$ (which is verified for all intervals); $\dot{E} = L/\epsilon_\mathrm{rad}$ is the total injected power, with the observed $\gamma$-ray luminosity $L$ given in Table~\ref{lorentz_factor}, radiative efficiency of the prompt $\epsilon_\mathrm{rad} = 0.1$; and magnetization parameter $\sigma \ll 1$ in the internal-shock scenario. 

The transparency condition $R_\mathrm{MeV} \geq R_\mathrm{ph}$ is satisfied in all intervals (Figure~\ref{lorentfactor_160625b}, panel C; Table~\ref{lorentz_factor}), with $R_\mathrm{MeV} \sim 10^{14}$~cm and $R_\mathrm{ph} \sim 10^{13}$~cm, confirming that the prompt emission is consistent with dissipation well above the photosphere, as expected in internal shock or some magnetic reconnection scenarios. We note, however, that satisfying the transparency condition does not exclude photospheric models, since sub-photospheric dissipation followed by radiative transfer could produce emission at comparable radii. However, an in-depth assessment of this alternative relies on detailed modeling of radiative transfer across the photosphere, which is beyond the scope of this work. 

\subsubsection*{Case (ii): Cutoff due to a natural break in the spectrum}

We now briefly consider an alternative interpretation, in which the observed high-energy cutoff reflects the intrinsic curvature of the inverse Compton spectrum rather than being caused by $\gamma\gamma$ pair production opacity. In this scenario, the cutoff arises naturally from the radiative process, and only a lower limit on the Lorentz factor can be derived. In each time interval, we have

\begin{equation}
    \Gamma_\mathrm{LL} = \max [\Gamma_{\gamma\gamma}(E_\mathrm{max})\,,\, \Gamma_\mathrm{tr}]
\end{equation}

where $E_\mathrm{max}$ is the maximum photon energy detected in each time interval (see Table \ref{lorentz_factor}), and $\Gamma_\mathrm{tr}$ is the lower limit of the Lorentz factor for transparency. Since $R_\mathrm{MeV} \geq R_\mathrm{ph}$ for all intervals, then $\Gamma_{\gamma\gamma}(E_\mathrm{max})\,\ge\, \Gamma_\mathrm{tr}$. The resulting lower limits are shown in Figure \ref{lorentfactor_160625b} (red symbols) and listed in Table \ref{lorentz_factor}. We find that these limits closely track the values of $\Gamma_{\gamma\gamma}(E_\mathrm{cutoff})$, while being always larger. 

The results presented in this section allow us to test the proposed phenomenological scenario discussed in Section \ref{interpretation}, in which the relation of time lags and spectral evolution for GRB~160625B is investigated in terms of the prompt emission of internal origin, which evolves hard to soft over time, and on the presence of the high-energy cutoff. 

\end{appendix}
\end{document}